\documentclass{iauc}
\usepackage{graphicx}

\title[SBF analysis pipeline for the astronomical community] 
{SAPAC: a SBF analysis pipeline for the astronomical community}

\author[Dunn \& Jerjen]   
{Laura P. Dunn
  \and Helmut Jerjen}

  \affiliation{RSAA, Mt Stromlo Observatory, ANU, Cotter Road, Weston ACT 2611 Australia \break 
    email: laurad@mso.anu.edu.au, jerjen@mso.anu.edu.au
  }

\pubyear{2005}
\volume{198} 
\pagerange{}
\date{?? and in revised form ??}
\jname{Near-Field Cosmology with Dwarf Elliptical Galaxies}
\editors{H. Jerjen \& B. Binggeli, eds.}
\begin{document}

\maketitle

\begin{abstract}
Large volumes of CCD imaging data that will become available from  
wide-field cameras at telescopes such as the CFHT, SUBARU, VST, or VISTA 
in the near future are highly suitable for systematic {\it distance surveys of early-type galaxies}  
using the Surface Brightness Fluctuation (SBF) method. For the efficient processing of such large data 
sets, we are developing the first semi-automatic SBF analysis pipeline named SAPAC. 
After a brief description of the SBF method we discuss the image quality needed for a 
successful distance measurement and give some background information on SAPAC.
\end{abstract}

\section{The SBF method in  a nutshell}
Employing the surface brightness fluctuation signal of unresolved stars in distant galaxies is an effective
and inexpensive new way to measure accurate distances to early-type (dwarf) galaxies. Unlike other extragalactic distance indicators (e.g. TRGB, RR Lyrae stars), this method does {\it not} require resolved stars therefore allowing distance measurements for early-type galaxies far beyond the practical limits of any of the classical distance indicators ($\sim$5\,Mpc). With Fourier analysis techniques, the SBF method quantifies the mean stellar flux per CCD pixel and rms variation due to Poisson noise across a designated area in a dwarf galaxy. 

Initially the SBF method was almost exclusively applied on nearby giant ellipticals and MW globular clusters (e.g. Tonry et al.~1989, 1994) but was found to work equally well with dwarf elliptical (dE) galaxies (e.g. Jerjen et al.~1998, 2000, 2001, 2004, and Rekola et al. 2005). As dE galaxies are by far 
the most numerous galaxy type at the current cosmological epoch, the SBF method in combination
with wide-field CCD imaging offers the opportunity for the first time to spatially locate dEs in vast numbers and thereby to map in 3D the densest environments of the local Universe (for first results see contributions by 
C\^ot\'e et al., Jerjen, Jordan et al., and Rekola et al.~in this volume). First SBF distances are published for dEs as distant as 15Mpc (using 2m ground-based telescopes) and 25Mpc (using 8m VLT+FORS and HST \& ACS).

\section{Analysis prerequisites}
The {\it minimal requirements} for the SBF analysis of an early-type galaxy are:
\smallskip

\begin{itemize}
\item Galaxy morphology: the light distribution of the stellar system must be radially symmetric and have 
minimal structure.  An overall elliptical shape of the galaxy is crucial as this is modelled 
and subtracted as part of the SBF analysis.
\smallskip

 \item{Photometry: calibrated CCD images are required in two photometric bands, e.g. ($B,R$) or ($g_{475}$, $z_{850}$), as the fluctuation magnitude shows a colour dependency.}
\smallskip

\item{Image quality: FWHM $ \leq  r_{\rm eff}['']/20$, where $r_{\rm eff}$ is the half-light radius of the galaxy.}
\smallskip

\item{Integration time: $t=$S/N$\cdot 10^{0.4\cdot(\mu_{\rm gal}-\mu_{\rm sky}+DM+\overline{M}-m_1)}$, 
where $\mu_{\rm gal}$ is the mean surface brightness of the galaxy, $\mu_{\rm sky}$ the surface brightness of the sky background, $DM$ the estimates distance modulus of the galaxy, $\overline{M}$ the fluctuation luminosity of the underlying stellar population, and $m_1$ the magnitude of a star 
providing 1 count/sec on the CCD detector at the telescope. }
\smallskip

\end{itemize}

\begin{figure}
\begin{centering}
 \includegraphics[height=0.47\textheight]{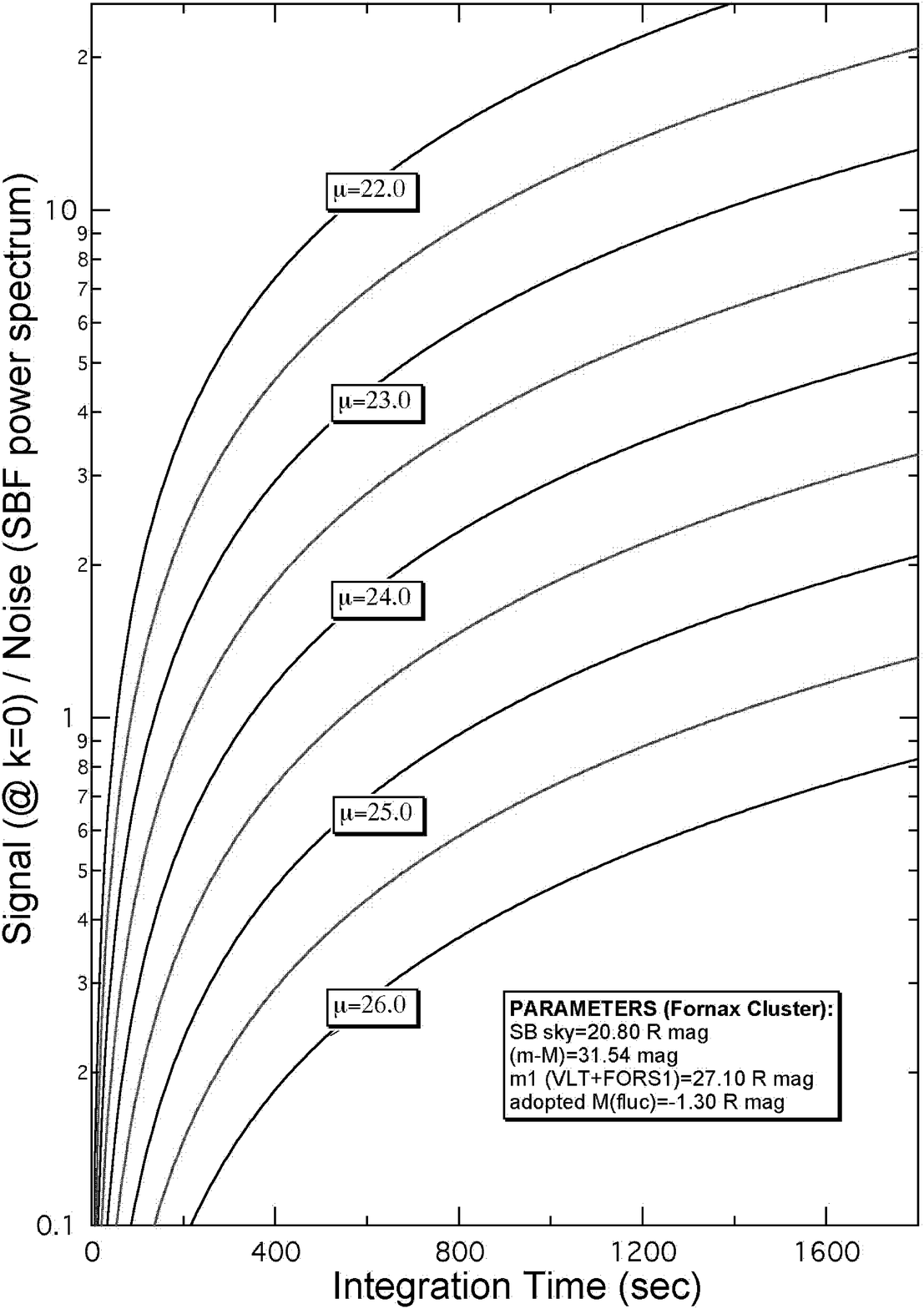}
  \caption{An illustration how the signal-to-noise in  the SBF  power spectrum increases 
  with length of exposure time and galaxy surface brightness at the distance of the Fornax 
  Cluster.}
  \label{fig1}
\end{centering}
\end{figure}

To give a general idea of these constraints, Fig.~\ref{fig1} illustrates the depth required for
an image of a dE at the distance of the Fornax cluster observed with VLT+FORS1. 
The SBF amplitude above the shot noise level (signal-to-noise) in the power spectrum is 
shown as a function of integration time and mean effective surface brightness of the galaxy. 
A SBF distance can be determined when the S/N is approximately 0.5, (see Fig.~8 in Rekola et al.~2005), but that depends largely on the image quality i.e. seeing.  For example, to 
achieve a S/N$\sim$2 in the galaxy power spectrum, the minimum exposure time required for a 
dE with a mean surface brightness of 25 mag\,arcsec$^{-2}$ is 1600s. 

It is interesting to note 
that this exposure time is by a factor of 20 shorter than the 32,000s of HST time spent by Harris et al.~(1998) 
to measure the TRGB distance of a dwarf elliptical at a similar distance.

\section{SBF Reduction Pipeline}
Previous SBF work has entailed individuals hand selecting regions in galaxy images for the analysis. 
To make the results as impartial as possible and data reduction more efficient we are developing  
a rapid, semi-automatic SBF analysis package named SAPAC that  can process large numbers of galaxies. 
SAPAC is a software package that carries out a semi-automatic SBF analysis of any early-type 
galaxy for which CCD data meets the requirements as discussed above.
For a detailed description of the fluctuation magnitude calibration and the individual reduction 
steps such as the modelling of the galaxy, foreground star removal, selection of SBF fields 
etc.~we refer the reader to Jerjen (2003).
SAPAC consists of Perl scripts using and IRAF module and uses a sophisticated graphical user interface, also written in 
Perl. The average processing time for 10 SBF fields in a galaxy  and measuring a distance is approximately 20 minutes. Initially we have concentrated the pipeline on $B$, $R$ images, but the 
implementation of calibration information for a wider range of commonly used filter sets  for
SBF work like $J,H,K$ of the SDSS $g, z$ filters is in process. 
\medskip

Potential users of SAPAC who are interested in testing this package for calculating accurate 
distances of early-type dwarfs are welcome to contact Laura Dunn.
This software package will be made available to the astronomical community soon. 

\begin{acknowledgments}
L.P.D would like to acknowledge partial financial support from the Astronomical Society of 
Australia, the International Astronomical Union, and Alex Rodgers Travel Scholarship.
\end{acknowledgments}

\end{document}